# SSVT: Self-Supervised Vision Transformer For Eye Disease Diagnosis Based On Fundus Images

Jiaqi Wang[1], Mengtian Kang[2], Yong Liu[1], Chi Zhang[2], Ying Liu[3], Shiming Li[2], Yue Qi[2], Wenjun Xu[2], Chenyu Tang[4], Edoardo Occhipinti[5], Mayinuer Yusufu[3], Ningli Wang[2], Weiling Bai[2], Shuo Gao[*1], Luigi G. Occhipinti[*4]

[1] School of Instrumentation and Optoelectronic Engineering, Beihang University, Beijing, China
[2] Beijing Tongren Hospital, Capital Medical University, Beijing, China
[3] Department of Surgery (Ophthalmology), The University of Melbourne, Melbourne, Australia
[4] Department of Engineering, University of Cambridge, Cambridge, UK
[5] Department of Computing, Imperial College London, UKRI Centre for Doctoral Training in AI for Healthcare, London, UK

*Abstract*— Machine learning-based fundus image diagnosis technologies trigger worldwide interest owing to their benefits such as reducing medical resource power and providing objective evaluation results. However, current methods are commonly based on supervised methods, bringing in a heavy workload to biomedical staff and hence suffering in expanding effective databases. To address this issue, in this article, we established a label-free method, named "SSVT", which can automatically analyze un-labeled fundus images and generate high evaluation accuracy of 97.0% of four main eye diseases based on six public datasets and two datasets collected by Beijing Tongren Hospital. The promising results showcased the effectiveness of the proposed unsupervised learning method, and the strong application potential in biomedical resource shortage regions to improve global eye health.

## I. Introduction

Eye diseases have become a global concern, strongly impacting patients' health conditions and regional economic growth. Among eye diseases, diabetic retinopathy (DR) [1], age-related macular degeneration (AMD) [2], glaucomatous optic neuropathy (GON) [3], and pathological myopia (PM) [4] are most widely observed, and they can lead to blindness when developing into late or even middle-stage [5]. Therefore, similar to many other diseases, broad-range screening for timely eye disease diagnosis is crucial as it is the basis for giving corresponding effective treatments [6]. Nevertheless, an undesired fact is that current eye disease diagnosis is mainly based on biomedical staff's subjective evaluation of fundus images, indicating that very different results may be yielded when various doctors are analyzing the same patient.

The above-explained issue not only results in ease of missing the optimal treatment windows for eye disease patients but also lacks good scalability, i.e. in developing countries, insufficient medical resources cannot support broad screening in this manner [7]. To address this issue, in recent years, artificial intelligent (AI) generated methods have started to release medical pressure and provide objective fundus image diagnosis results. To date, promising results have been reported [8]–[15], and state-of-the-art techniques can detect eye diseases such as DR, AMD, GON, and others, via training with labeled fundus images. However, current methodologies frequently require manual annotation by medical practitioners, involving the meticulous task of categorizing fundus images with accurate disease classifications or regions of interest. Additionally, these methodologies primarily focus on specific eye diseases, limiting their adaptability in more extensive screening scenarios.

In response to these challenges, this article introduces the Self-Supervised Vision Transform (SSVT) network for the diagnosis of four major eye diseases, DR, AMD, GON, and PM, along with the assessment of disease severity through fundus images, within a diagnostic framework process shown in Fig. 1(a). The key innovations presented in this work encompass:

(1) Establishment of a self-supervised learning framework to mitigate data acquisition costs and enhance generalizability to diverse eye diseases.

(2) The extraction of semantic vectors from fundus images utilizing the self-attention mechanism, enhancing the model's efficacy in capturing global information within the images.

These advancements contribute significantly to the evolving landscape of automated eye disease diagnosis, offering a promising avenue for scalable, accurate, and efficient screening while alleviating the constraints associated with manual annotation processes. SSVT exhibits superior performance across six publicly available datasets and two self-collected datasets with an average area under the receiver operating characteristic curve (AUC) of 93.5% and an average (ACC) of 97.0%.

## II. Materials and methods

### A. Dataset and labeling

101,064 color fundus photographs from six publicly available datasets (EyePACS [16], Messidor [17], Messidor-2 [18], and APTOS-2019 [19] for DR, Ichallenge-AMD [20] for AMD, REFUGE for GON, and Ichallenge-PM [21] for PM) and two self-collected external datasets from Beijing Tongren Hospital (a GON dataset with 1,919 images and a PM dataset with 2,237 images) were collected to develop SSVT.

Image labelling enlisted five junior ophthalmologists, each with a minimum of 3 years of clinical experience, with validation by three senior retinal specialists having over 10 years of expertise. All fundus images were labelled based on lesion severity for DR, AMD, GON, and PM. DR was classified into five classes (no evidence, mild, moderate, severe, and proliferative), while AMD, GON, and PM were categorized as either not apparent or apparent.

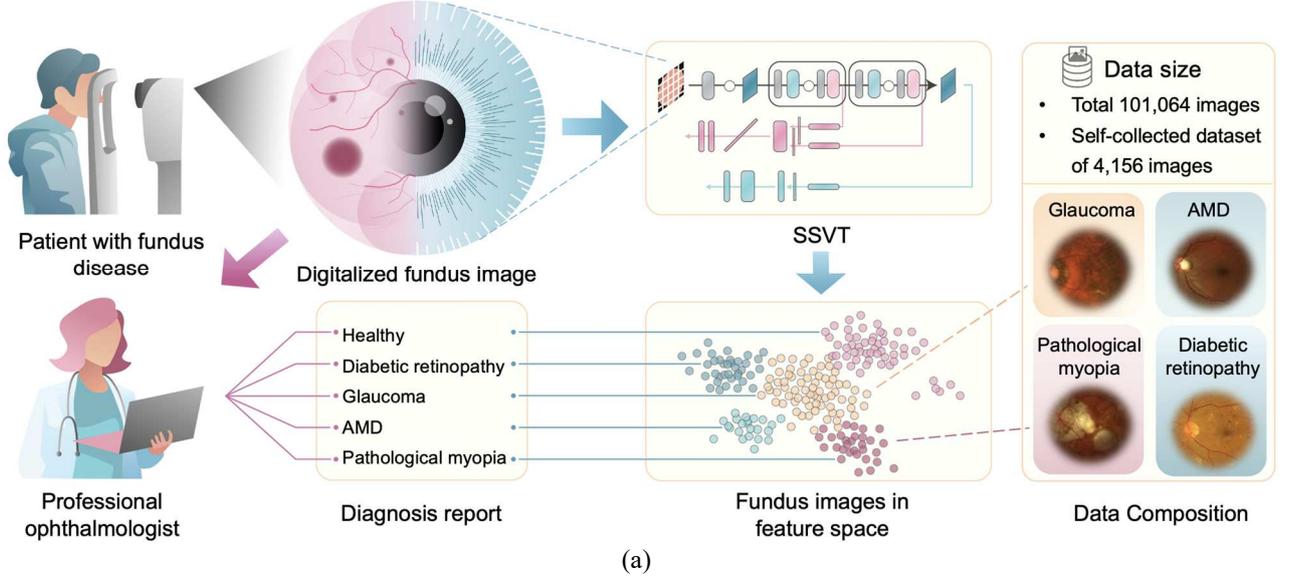

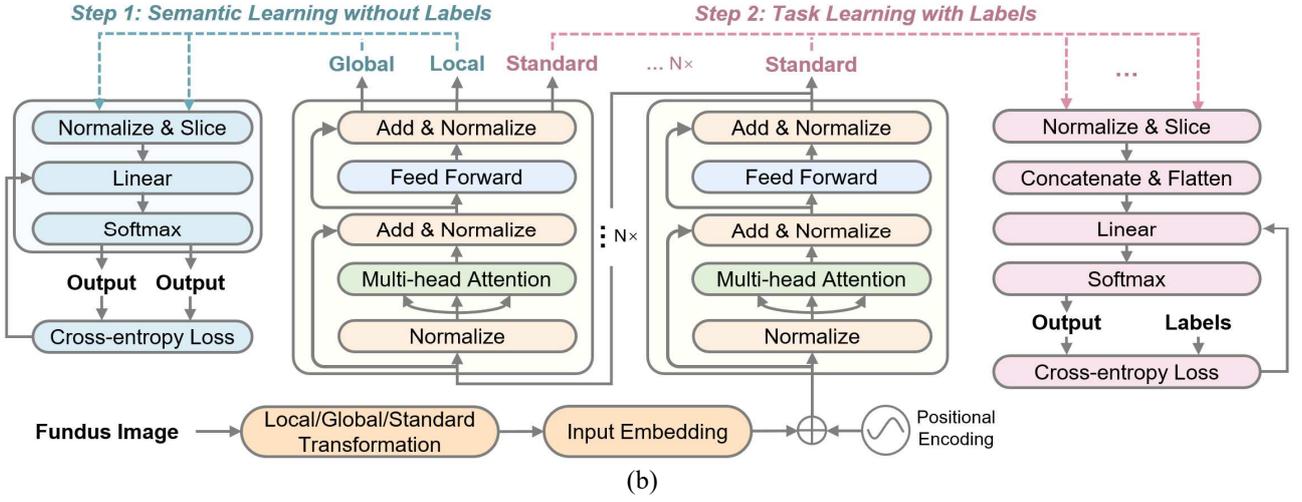

Fig. 1. (a) Description of the methodology. (b) The network structure diagram of the SSVT.

*B. Algorithm development*

SSVT constructed upon the Data-efficient image transformers Base resolution (Deit-B) architecture [22] was implemented in two stages, as shown in Fig. 1. (b).

Initially, 88,703 normalized fundus images from EyePACS dataset were utilized for self-supervised learning, generating a high-dimensional vector embedding with semantic information. Specifically, each image was transformed into global and local replicas, which were entered into the global Deit-B model and local Deit-B model respectively after the random horizontal flip and grayscale transformation, as described in Eq. (1) and Eq. (2)

$$X_g = \bigcup_{x \in X}\{F_g(x,\varphi_i), i = 1, \ldots, n_g\} \quad (1)$$

$$X_l = \bigcup_{x \in X}\{F_l(x,\varphi_i), i = 1, \ldots, n_l\} \quad (2)$$

where $X_{g/l}$ represents global or local crop sets (*g* for global, *l* for local) with predefined crop numbers $n_{g/l}$, $F_{g/l}(x,\varphi_i)$ is the global or local crop operations, where $\varphi_i$ represents subsequent transformations including random crop and flip.

The global and local Deit-B models share identical structures, with the gradient function excluded in the global Deit-B model. Both models output high-dimensional feature vectors, and the cross-entropy loss between them is calculated after Softmax operation, as described in Eq. (3)-(5)

$$P_t^i(x_g, \theta_t) = \frac{\exp(G_t^i(x_g, \theta_t)/\tau_t)}{\sum_{k=1}^K \exp(G_t^k(x_g, \theta_t)/\tau_t)} \quad (3)$$

$$P_s^i(x_l, \theta_s) = \frac{\exp(G_s^i(x_l, \theta_s)/\tau_s)}{\sum_{k=1}^K \exp(G_s^k(x_l, \theta_s)/\tau_s)} \quad (4)$$

$$H(\alpha, \beta) = -\alpha \log \beta \quad (5)$$

where $G_{t/s}^i(x_{g/l},\theta_{t/s})$ means the *i*-th class output among *K* classes of global or local crop sets (*t* for global, *s* for local), $P_{t/s}^i(x_{g/l},\theta_{t/s})$ represents the probability of the *i*-th class output calculated via Softmax operation with temperature $\tau$, $x_g$ corresponds to the input global crop, $\theta_{t/s}$ is the global/local model's weights, and $H(\alpha, \beta)$ is the cross-entropy loss. The loss updates the weight parameters of the local model by back-propagation through Eq. (6). Conversely, the global model was updated by an exponential moving average of the historical weights of the local model.

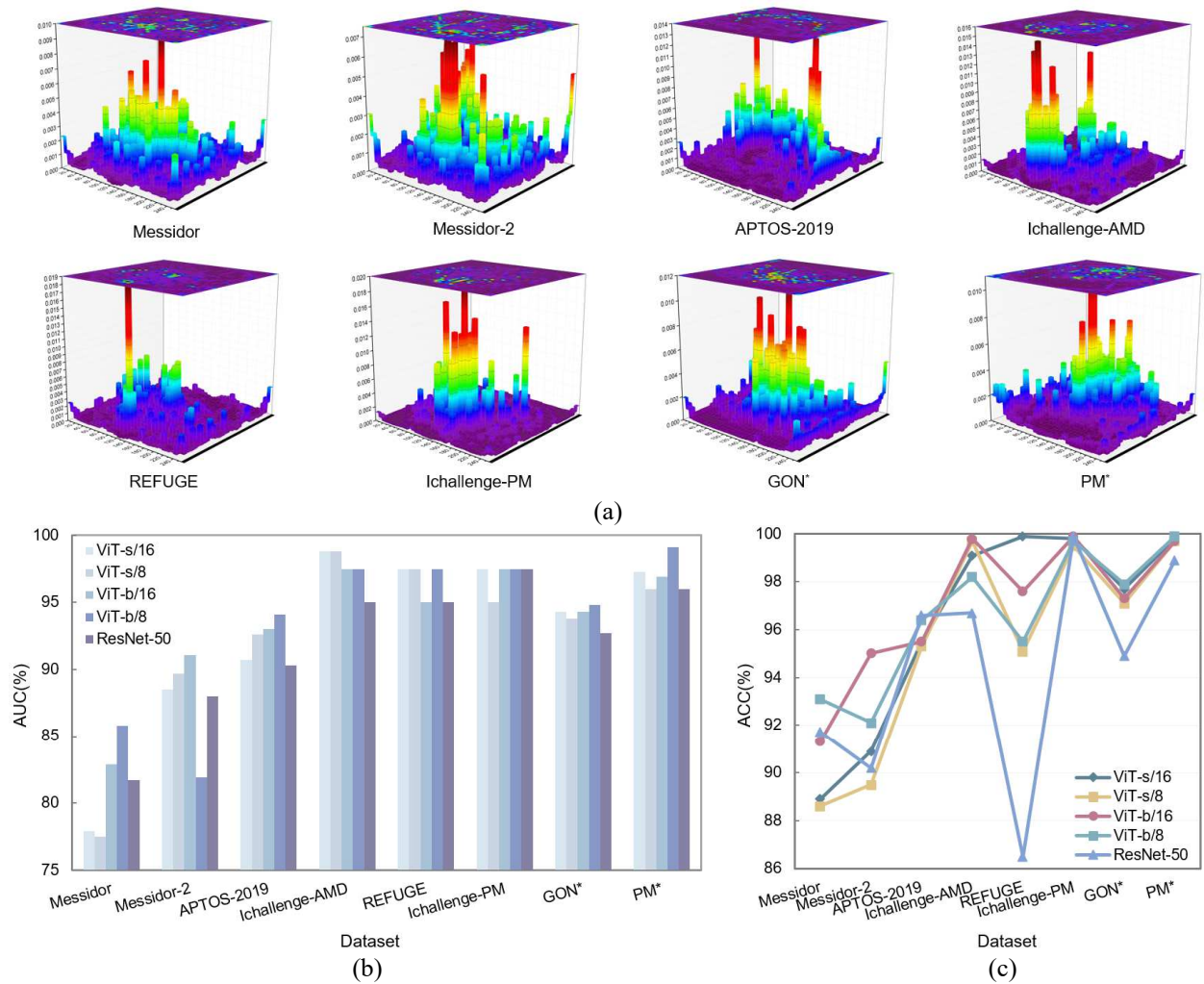

**Fig. 2.** (a) Fundus image data representation using self-supervised machine learning. (b) AUC comparison of different models. (c) ACC comparison of different models. Privately collected datasets are marked with *.

$$\min_{\theta_s} \sum_{x_g^i \in X_g, x_l^i \in X_l} H(P_t(x_g^i, \theta_t), P_s(x_l^i, \theta_s))$$
$$\text{subject to } \theta_t^{epoch} = \lambda \theta_t^{epoch-1} + (1-\lambda)\theta_s^{epoch-1}$$
(6)

In the second stage, eye diseases were determined using datasets such as Messidor-2, Ichallenge-AMD, REFUGE, and Ichallenge-PM. The high-dimensional feature vectors obtained from the global model, trained through self-supervised learning, were utilized to construct a linear classifier, a single-layer neural network, for eye disease classification. The ratio of training, validation, and test sets is 0.6:0.2:0.2. Stratified partitioning was performed to ensure an even distribution of fundus image data across classes.

*C. Evaluation protocol*

AUC and ACC were employed as evaluation metrics for the performance of the deep learning system. In instances of multiple classes of lesions in diabetic retinopathy (DR), separate AUC values were computed for each category, and the average value was derived as the conclusive result.

### III. EXPERIMENTS AND RESULTS

*A. Data characteristics*

The self-supervised learning component of SSVT was established with 88,702 unlabeled fundus images. Subsequently the linear classifier for disease diagnosis was developed employing 9,890 labeled fundus images covering DR, glaucoma, age-related macular degeneration, and pathological myopia. The evaluation of SSVT's performance involved 1,641 external validation images collected from Beijing Tongren Hospital, comprising participants undergoing annual health checks.

*B. Performance comparison*

The SSVT was systematically assessed across six public eye disease datasets and two external datasets. According to the model complexity and patch size, 4 models were designed based on the small structure Vision Transformer model (ViT) [23] under 16×16 and 8×8 patch size (ViT-s/16 and ViT-S/8), as well as the big structure Vision Transformer Based Resolution under 16×16 and 8×8 patch size (ViT-b/16 and ViT-b/8). The semantic vectors extracted by ViT-b/16 are shown in Fig. 2. (a). Additionally, a Residual Network (ResNet) was employed to facilitate model performance comparison. As depicted in Fig. 2. (b)-(c), the ViT-s/16-based SSVT exhibited an average AUC of 92.8% and an average ACC of 96.5% across all datasets. The ViT-s/8-based SSVT achieved an average AUC of 92.6% and an average ACC of 95.6% across all datasets. For ViT-b/16-based SSVT, an average AUC of 93.5% and an average ACC of 97.0% were achieved across all datasets, while ViT-b/8-based SSVT secured an average AUC of 93.5% and an average ACC of 96.6%. In comparison, ResNet demonstrated an average AUC of 92.0% and an average ACC of 94.4% across all datasets.

## IV. Discussion

In this study, SSVT, an innovative label-free self-supervised vision transformer network was designed for automatic eye disease diagnosis. The SSVTs based on ViT-s/16, ViT-s/8, ViT-b/16, and ViT-b/8 demonstrated state-of-the-art performance in 4 disease classifications and grades across four public and two external eye disease datasets. The results indicated that patch size and model complexity had slight effects on overall performance. Notably, SSVT based on ViT-b/16 emerged as a standout performer, highlighting the efficacy of employing larger patch sizes in the Vision Transformer (ViT) framework within the context of self-supervised learning for eye disease classification and grade. Furthermore, a comparative analysis with Residual Network (ResNet) was conducted. The ViT-based methods achieved 1.6%-2.3% higher ACC than ResNet. The results suggested that self-attention mechanism-based architecture possesses enhanced adaptability in global feature learning tasks.

## V. Conclusions

Machine learning is demonstrating its power in many healthcare applications. In this article, our SSVT model exhibits a high eye diseases diagnosis accuracy of 97.0%, which is comparable or even beyond human expert's accuracy. Attention to privacy and data sharing policies of different biomedical institutions in their respective countries mean that wider deployment of the proposed method relies on the implementation of data protection measures and open access schemes across institutions and countries. We are confident that this work stimulates further research and adoption of the proposed model and knowledge base, contributing to innovative and effective health care solutions for the prevention and management of eye diseases.

## VI. Compliance with Ethical Standards

This study was approved by the Ethics Committee of Beijing Tongren Hospital, Capital Medical University, and adhered to the tenets of the Declaration of Helsinki. Informed consent was not obtained from patients due to the anonymity and retrospective nature of the study.

## Acknowledgments

This work was supported in part by the National Natural Science Foundation under Grant 62171014, National Natural Science Foundation of China 82201244, Natural Science Foundation of Beijing M22019, Beijing Hospitals Authority Innovation Studio of Young Staff Funding Support 202106 and from the EPSRC, under grants EP/K03099X/1 and EP/W024284/1. E.O. was supported by UKRI Centre for Doctoral Training in AI for Healthcare grant EP/S023283/1.